\documentstyle[aps,prb,amsfonts,multicol,epsf]{revtex} 
\begin{document}
\draft
\title{On the $3n+l$ Quantum Number in the Cluster Problem}
\author{Erik Koch}
\address{Max-Planck-Institut f\"ur Festk\"orperforschung, D-70569 Stuttgart}
\date{20 November 1995}
\maketitle 

\vspace{3ex}
\begin{abstract}
It has recently been suggested that an exactly solvable problem 
characterized by a new quantum number may underlie the electronic 
shell structure observed in the mass spectra of medium-sized sodium 
clusters. We investigate whether the conjectured quantum number $3n+l$ 
bears a similarity to the quantum numbers $n+l$ and $2n+l$, which 
characterize the hydrogen problem and the isotropic harmonic oscillator 
in three dimensions.
\end{abstract}
\pacs{36.40.Cg, 71.20.-b, 03.65.Fd\hfill cond-mat/9606050}

\begin{multicols}{2}
\section{Introduction}

The electronic shell structure observed in mass spectra of metal
clusters can be rather well described by an independent electron model.
Treating the valence electrons of the cluster as an ideal Fermi gas
moving in an external potential $V(\vec{r}\,)$, a shell structure arises
from the discreteness of energy levels for particles confined to a
finite volume. This approach forms the basis for the semiclassical
description of the electronic shell and supershell structure.
\cite{nishioka90,bennemannprl,gutzwiller70,BaBlo3,Lerme93a,BrackRev} 
Suitable model potentials can be obtained from self-consistent density
functional calculations.\cite{BrackRev,ekardt84,GenzkenPRL} 
It can be verified that the electronic shell
structure may be determined from the spectrum of a one-particle
Hamiltonian, provided the $V(\vec{r}\,)$ is chosen sufficiently close
to the self-consistent potential.\cite{HarrisArgument}

For small sodium clusters of 8 up to 92 atoms it has been shown that
the shell structure arises from successively filling the
$2(2l+1)$-fold degenerate energy levels $\epsilon_{n,l}$ of a
spherically symmetric potential. \cite{Knight84} To uniquely
characterize the energy levels for such potentials we use the number
$n$ of nodes in the radial wave function, while $l$ denotes the
angular momentum quantum number.\cite{nqm} Going to larger clusters reduces the 
influence of a single energy level on the shell structure.
Nevertheless even for clusters of several thousand atoms
electronic shell structure has been observed. \cite{prl90,Bjornh90,cpl91}
This suggests that the
energy levels bunch together in groups of (nearly) constant energy.
Actually, the experimental results for sodium clusters of up to
approximately 1500 atoms\cite{cpl90,jpc91,zpd91} indicate that these 
bunches may be formed by degenerate energy levels $\epsilon_{n,l}$ 
corresponding to definite values of $3n+l$. It has been pointed 
out that quantities of the type $\alpha n+l$ characterize two
exactly solvable quantum mechanical problems: the hydrogen atom, in which 
levels with the same $n+l$ are degenerate, and the isotropic
harmonic oscillator, where the degeneracy is described by the quantum
number $2n+l$. With this analogy in mind it has been 
conjectured\cite{cpl90,jpc91,zpd91} that there
might be a set of solutions of the radial Schr\"odinger equation
displaying the degeneracy described by the quantum number $3n+l$.

At first sight a theorem from classical mechanics seems to imply that
there is no such radial potential. The theorem states that only for central
potentials of the form $-1/r$ and $r^2$ {\em all} bounded orbits are 
closed.\cite{Arnold}
Assuming that this property is caused by a dynamical symmetry of the
system which uniquely corresponds to a symmetry of the quantized system,
this would imply that only these two problems exhibit ``accidental''
degeneracies. Unfortunately the situation is not that simple. While it 
is known that the degeneracies for the quantum mechanical Kepler problem
and the harmonic oscillator are a consequence of the dynamical symmetries 
${\frak so}(4)$ and ${\frak su}(3)$, respectively, it turns out that {\em all}
classical systems involving 3 degrees of freedom automatically possess both
${\frak so}(4)$ {\em and} ${\frak su}(3)$ symmetry.\cite{ClassSymm}
To our knowledge, the question whether the hydrogen atom and the harmonic
oscillator are the {\em only} radially symmetric quantum mechanical systems 
possessing a dynamical symmetry has not been settled so far. A discussion
of near degeneracies of the type $\alpha n + \beta l$ in the context of
semiclassical quantization has been given by Bohr and Mottelson.\cite{BoMo}

Besides being interesting in its own right, identifying an exactly solvable 
problem possessing a quantum number $3n+l$ would be important for two reasons. 
First, it would provide us with the optimal basis set for self-consistent 
computations of the total energy of metal clusters. Second, it would give the 
ideal starting point for treating realistic cluster potentials by
perturbation theory.

In the present paper, we investigate whether the number $3n+l$ is
analogous to the quantum numbers $n+l$ and $2n+l$.
First we argue that a degeneracy described by the
quantum number $3n+l$ cannot be caused by a dynamical symmetry
related to a classical Lie algebra. Next we show that in three
dimensions there are exactly two potentials fulfilling the
shape-invariance condition of supersymmetric quantum mechanics for
spherically symmetric problems, namely the Coulomb potential and the
harmonic oscillator. Finally we numerically optimize radial potentials
to exhibit the $3n+l$-degeneracy. We find, within the limits of our
calculations, that there is no exactly solvable
`cluster potential' with $3n+l$-degeneracy. We therefore 
conclude that $3n+l$ is not a proper quantum number as $n+l$ and $2n+l$ 
are.

\noindent
\begin{minipage}{3.375in}
\begin{table}
  \begin{tabular}{lcc}
    system & quantum number & symmetry\\
    \tableline 
    hydrogen atom       & $\; n+l\;$ & ${\frak so}(4)$ \\[1ex]
    harmonic oscillator & $\;2n+l\;$ & ${\frak su}(3)$ \\[1ex]
    cluster             & $\;3n+l\;$ & {\large\bf ?} \\
  \end{tabular}
  \vspace{1ex}
  \caption[]{\label{analogy}
     Possible analogy between systems with a dynamical symmetry and
     a conjectured `cluster problem', being characterized by a quantum
     number $3n+l$.}
\end{table}
\end{minipage}
 
\section{Classical Symmetry}

A quantum mechanical system is said to exhibit a symmetry if its
Hamiltonian is invariant with respect to the operations of some Lie
algebra. A well-known example is spherical symmetry: the Hamiltonian
$H$ of a radially symmetric system commutes with the angular momentum
operators:
\begin{equation}\label{commute}
  [L_i,H] = 0 .
\end{equation}
Since $\{L_x, L_y, L_z\}$ forms a Lie algebra, namely the angular
momentum algebra ${\frak su}(2)$, the Hamiltonian is said to be
invariant under ${\frak su}(2)$. As an immediate consequence of
(\ref{commute}) the eigenspaces ${\cal H}_{\epsilon_{n,l}}$ of the
Hamiltonian are invariant under the action of the $L_i$, i.e.\ they
are representation spaces of the symmetry algebra. For a generic
radial potential, the ${\cal H}_{\epsilon_{n,l}}$ will be irreducible
representation spaces of the angular momentum algebra. For these
systems ${\frak su}(2)$ is the full symmetry.

There are, however, special cases in which the eigen\-spaces of the Hamiltonian
are reducible. Examples are the hydrogen atom and the isotropic
harmonic oscillator. For the former, energy levels $\epsilon_{n,l}$
having the quantum number $n+l$ in common are degenerate, for the
latter the respective quantum number is $2n+l$.  It is well known that
these degeneracies are not accidental, rather they are the consequence
of a symmetry higher than spherical. For the hydrogen atom this symmetry is
described by the algebra ${\frak so}(4)$ while for the harmonic
oscillator the symmetry algebra is the ${\frak su}(3)$. \cite{Wybourne}
Having identified $3n+l$ as a candidate for a new quantum number, the
question arises as to whether the corresponding degeneracy is also due to
some hidden symmetry. 

To answer this question, we exploit the fact that eigenspaces of a Hamiltonian
are representation spaces of its symmetry algebra. In particular, the
dimensions of these spaces must coincide. Fortunately, representation
theory \cite{Humphreys,FultonHarris} provides us with all the
information needed. The irreducible representations of the classical
Lie algebras (${\frak su}(k+1)$, ${\frak so}(2k+1)$, ${\frak sp}(2k)$,
and ${\frak so}(2k)$, corresponding to the algebras $A_k$, $B_k$,
$C_k$, and $D_k$ in the complete classification scheme of all semisimple Lie
algebras) can be labeled using the Cartan subalgebra. For a Lie
algebra of rank $k$ this is a set of $k$ commuting operators. They can
be chosen such that they have integer eigenvalues. For any given
representation one can find a basis which diagonalizes the operators
in the Cartan subalgebra. Each basis vector is then characterized by
the corresponding integer vector of eigenvalues $(m_1,\ldots,m_k)$,
its `weight'. Furthermore each irreducible representation is uniquely
determined by the highest weight of its basis vectors according to
lexicographical ordering. The dimension of the irreducible
representation with highest weight $(n_1,\ldots,n_k)$ is given by
Weyl's dimensionality formula
\begin{equation}
  {\rm dim}\Big(\Gamma_{\cal L}(n_1,\ldots,n_k)\Big) = 
    \prod_{\alpha\in\Delta_+} 
       \left({ \sum\limits_{j=1}^k c_j^{(\alpha)}\,n_j \over
               \sum\limits_{j=1}^k c_j^{(\alpha)}}
                                    + 1 \right) .
\end{equation}
The constants $c_j^{(\alpha)}$ and the set $\Delta_+$ of positive roots
only depend on the type of the Lie algebra, not on any
particular representation. Hence for a given classical Lie algebra,
the dimensions of the irreducible representations are given by a
polynomial in the highest weights. As an example we write down
explicit expressions for the dimensions of the classical algebras of
rank two:
\begin{eqnarray}
 {\rm dim}\Big(\Gamma_{A_2}(n_1,n_2)\Big) \label{dimA2}
    &=&{(n_1+1)(n_1+n_2+2)(n_2+1)\over2}, \\
 {\rm dim}\Big(\Gamma_{B_2}(n_1,n_2)\Big) && \\ &&\hspace*{-14ex} 
     = {(n_1+1)(n_1+n_2+2)(2n_1+n_2+3)(n_2+1)\over6}, \nonumber\\
 {\rm dim}\Big(\Gamma_{D_2}(n_1,n_2)\Big) \label{dimD2}
    &=&(n_1+1)(n_2+1).
\end{eqnarray}
These have to match the dimensions of the eigenspaces of the
Hamiltonian. Given the quantum number $N$ associated with the degeneracy of
energy levels $\epsilon_{n,l}$, the orders $g_N$ of the degeneracies
can be readily computed. For the quantum number $N_1 = n+l$ we obtain
\begin{equation}
  g_{N_1} = (N_1+1)^2 ;
\end{equation} 
$N_2 = 2n+l$ leads to
\begin{equation}
  g_{N_2} = {(N_2+1)(N_2+2)\over 2},
\end{equation}
and $N_3 = 3n+l$ to 
\begin{equation}
  g_{N_3} = \left\{
   \begin{array}{cc}
    {1\over3}   (N_3+2)^2  ,&\mbox{ for $N_3\;{\bf mod}\;3 = 1$},\\[1.5ex]
    {1\over3}(N_3+1)(N_3+3),&\mbox{ otherwise}.
   \end{array}\right. 
\end{equation}
Through comparing with expressions (\ref{dimA2})-(\ref{dimD2}), we find
\begin{eqnarray}
  g_{N_1} &=& {\rm dim}\Big(\Gamma_{D_2}(N_1,N_1)\Big) ,\\
  g_{N_2} &=& {\rm dim}\Big(\Gamma_{A_2}(N_2, 0 )\Big) ,
\end{eqnarray}
indicating that the eigenspaces of the Hamiltonian for the hydrogen
atom might be irreducible representation spaces of the Lie algebra
$D_2 \equiv {\frak so}(4)$ with highest weights $(n_1,n_2) = (N_1,N_1)$
--- as indeed it is. Similarly for the harmonic oscillator the eigenspaces 
could be the irreducible representation spaces of $A_2 \equiv {\frak su}(3)$ 
with highest weights $(n_1,n_2) = (N_2,0)$ --- again, as indeed it is. 
For the quantum number $3n+l$ the situation is fundamentally different. 
The degeneracies of the eigenspaces are no longer given by a closed formula, 
i.e.\ there is no simple relation between the quantum number and the 
highest weights of some Lie algebra. 
If there existed a quantum mechanical operator corresponding to the quantum 
number, which would be a Casimir operator of the symmetry algebra, then such a 
simple relation should exist, since the eigenvalues of a Casimir operator are 
given by rational functions in the highest weights.\cite{Casimirs}
Thus, we cannot find a classical Lie algebra that could cause a degeneracy 
corresponding to the quantum number $3n+l$. The same reasoning holds for 
quantum numbers $\alpha n + l$ with $\alpha > 3$.

\section{Supersymmetry}

An elegant approach to exactly solvable problems is provided by
supersymmetric quantum mechanics. \cite{SuSyRep,Sukumar2} The basic
idea of the method is that a given Hamiltonian
\begin{equation}
  H^- \equiv -{d^2\over dx^2} + V^-(x)
\end{equation}
can be factorized if the energy of the ground state $\Psi_0^-(x)$ is set
to zero:
\begin{equation}
  H^- = Q^+ Q^- \;\mbox{, where } 
  Q^\pm \equiv \left(\mp{d\over dx} - {(\Psi_0^-)'\over\Psi_0^-}\right) .
\end{equation}
Here the prime denotes the derivative with respect to $x$.
Reversing the order in the factorization, a new Hamiltonian $H^+ \equiv Q^-
Q^+$ can be defined. It is called the supersymmetric partner of $H^-$.
Obviously the $Q^\pm$ intertwine the Hamiltonians:
\begin{eqnarray}
  H^- Q^+ &=& Q^+ H^+ ,\\
  H^+ Q^- &=& Q^- H^- .
\end{eqnarray}
Thus, except for the ground state of $H^-$, which is annihilated by
$Q^-$, the spectra of the Hamiltonians $H^\mp$ are identical. This
situation is illustrated in Fig.~\ref{SuSy.epsi}. 

The potentials corresponding to the Hamiltonians $H^\mp$ can be
derived from the superpotential 
\begin{equation}\label{superpot}
W(x) \equiv -{(\Psi_0^-)' \over \Psi_0^-}
\end{equation}
via
\begin{equation}\label{Vsuper}
V^\mp(x) = \Big(W(x)^2 \mp W'(x)\Big) .
\end{equation}
For a generic Hamiltonian the potential $V^-(x)$ and its
supersymmetric partner $V^+(x)$ will be of different shape. 

\noindent
\begin{minipage}{3.375in}
\begin{figure}
  \centerline{\epsffile{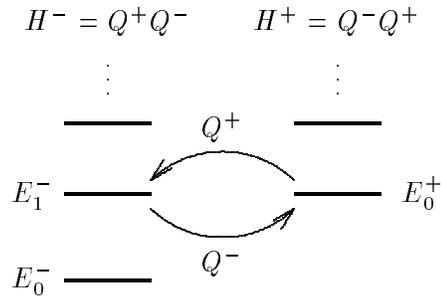}}
  \vspace{3ex}
  \caption[]{\label{SuSy.epsi}
    Spectra of supersymmetric partner Hamiltonians $H^-$ and $H^+$
    (non broken supersymmetry).}
\end{figure}
\end{minipage}

\vspace{2ex}
\noindent
If, however,
they only differ in some parameters and a constant offset, i.e.\ if
\begin{equation}\label{shapeinvar}
  V^+(a_0;x) = V^-(a_1,x) + R(a_0),
\end{equation}
the corresponding Hamiltonian is exactly solvable. \cite{Gendenshtein}
In what follows we shall refer to the expression in equation (\ref{shapeinvar})
as the shape-invariance condition.
In such a case, starting from the potential $V^-(a_0;x)$, a whole
sequence 
\begin{equation}
  V^-(a_k;x) \stackrel{\rm SuSy}{\longrightarrow} 
  V^+(a_k;x) = V^-(a_{k+1};x) + R(a_k)
\end{equation}
of supersymmetric partner potentials can be constructed. From the
knowledge of the ground-state energy in the potential $V^-(a;x)$ as a
function of the parameters $a$, the whole spectrum of the initial
Hamiltonian $H^-$ can be obtained:
\begin{equation}
  E_n^- = \sum_{k=0}^{n-1} R(a_k) .
\end{equation}

We now investigate whether a radial problem with degeneracies
according to the quantum number $3n+l$ can be constructed using the
above ideas. More specifically, we are looking for any potential $U(r)$
such that energy levels with different angular momentum $l$ are
degenerate. For the radial potentials
\begin{equation}\label{radpot}
  V(l;r) \equiv U(r) + {l(l+1)\over r^2}
\end{equation}
the shape-invariance condition (\ref{shapeinvar}) reads
\begin{equation}\label{radshapeinv}
  V^+(l;r) = V^-(\tilde{l};r) + R(l).
\end{equation}
Using (\ref{radpot}) this can be rewritten as 
\begin{equation}\label{radshapeinvII}
  V^+(l;r) = V^-(l;r) + {2\lambda_l\over r^2} + 2\beta_l,
\end{equation}
where we have introduced the notations
\begin{equation}
  2\,\lambda_l = l(l+1) - \tilde{l}(\tilde{l}+1) \quad\mbox{and}\quad
  2\,\beta_l   = R(l) .
\end{equation}
Inserting (\ref{Vsuper}) into (\ref{radshapeinvII}), we obtain a differential 
equation for the superpotential $W(l;r)$, the general solution of which is
\begin{equation}\label{Wansatz}
  W(l;r) = -{\lambda_l\over r} + \alpha_l + \beta_l\,r .
\end{equation}
The corresponding potentials are
\begin{eqnarray}\label{Vansatz}
  V^\mp(l;r) =&&   {\lambda_l(\lambda_l\mp1)\over r^2}
                 - {2\lambda_l\alpha_l\over r}          \\
              && - (2\lambda_l\pm1)\beta_l
                 + (\alpha_l+\beta_l r)^2               \nonumber .
\end{eqnarray}
The choices $\lambda_l=l+1$ and $\lambda_l=-l$ will make $V^-(l;r)$ a 
radial potential with angular momentum $l$.
To ensure that we can normalize the ground state, the parameters in the
above equations must satisfy 
\begin{equation}
 \lambda_l > -{1\over2} ,\quad\mbox{and}\quad
 \left\{\begin{array}{c}
   \beta_l > 0 , \\ 
   \mbox{or}\\
   \beta_l = 0 , \quad \alpha_l > 0 .
 \end{array}\right. 
\end{equation}
We thus have to choose $\lambda_l=l+1 > -1/2$. Inserting this into equation
(\ref{Vansatz}) yields
\begin{eqnarray}
  V^+(l;r) &-& V^-(l+1,r) \\
     =&& \Big\{ (\alpha_l    +\beta_l     r)^2        \nonumber
              -(\alpha_{l+1}+\beta_{l+1} r)^2 \Big\} \\[1ex]
     &-& \left\{ {2(l+1)\alpha_l    \over r}          \nonumber
                -{2(l+2)\alpha_{l+1}\over r} \right\} \\[1ex]
     &+& \Big\{ -(2l+1)\beta_l            \nonumber    
                +(2l+5)\beta_{l+1} \Big\} .
\end{eqnarray}
To be consistent with (\ref{shapeinvar}), we require the right-hand
side of the above equation to be constant. By comparing the coefficients, we
find exactly two solutions of the shape-invariance condition:
\begin{eqnarray}
 {\alpha_{l+1}\over\alpha_l} = {l+1\over l+2} 
              &\mbox{and}& \beta_l=0 ,\\
              &\mbox{or}& \nonumber  \\
 \alpha_l = 0 &\quad\mbox{and}\quad& \beta_l = const.
\end{eqnarray}
The first solution corresponds to the Hamiltonian for the hydrogen
atom. Setting $\alpha_l = q/(2(l+1))$ we obtain the energy spectrum
\begin{equation}\label{hydrospect}
  \epsilon_{n,l} = E(l)_n^- = -{q^2\over4(n+l+1)^2}
\end{equation}
pertaining to the Coulomb potential $V(r)=-q/r$. Writing the constant in the
second solution as $\beta_l=\omega/2$ we find the spectrum of the
harmonic oscillator $V(r)=\omega^2 r^2 / 4$:
\begin{equation}\label{harmospect}
  \epsilon_{n,l} = E(l)_n^- = (2n+l+3/2)\omega .
\end{equation}
Obviously, the spectra of these systems are characterized by
degeneracies according to the quantum numbers $n+l$ and $2n+l$. Thus we
have shown that there exist exactly two physically acceptable {\em radial} 
potentials which fulfill the shape-invariance condition (\ref{radshapeinv}), 
namely those associated with the quantum numbers $n+l$ and $2n+l$. 
It thus appears that these two quantum numbers are unique. 

\section{Numerical optimization}

Clearly, arguments of the type given above can only exclude the
existence of a system with quantum number $3n+l$ under the assumption 
of some specific mechanism (e.g.,~classical symmetry or supersymmetry) 
that would cause the degeneracy. Thus, by such reasoning it is not possible 
to give a general proof that such a system cannot exist.
Furthermore, it is conceivable that $3n+l$ is only an
approximative quantum number, i.e.\ that levels of quantum number
$3n+l$ are not exactly but merely near-degenerate. Therefore, we will now
approach the problem at hand from a completely different direction. 

We try to construct cluster-potentials exhibiting degeneracies
according to $3n+l$ by numerical optimization. The basic idea is to
transform the continuous eigenvalue problem for the Hamiltonian
\begin{equation}
  H = -{\hbar^2\over2m}\Delta + V(r)
\end{equation}
into a finite dimensional problem by discretizing the Schr\"o\-dinger
equation. Discretizing the potential
\begin{equation}
  \Big\{V(r)\;\Big|\,r\in [0,\infty)\Big\} 
 \mapsto \Big\{V_i\;\Big|\,V_i = V(r_i)\Big\} ,
\end{equation}
the eigengenvalues can be considered as {\em functions}
of a finite number of potential parameters $V_i$ instead of being 
{\em functionals} of the potential:
\begin{equation}\label{Ediscrete}
  \epsilon_{n,l}\Big[V(r)\Big] \approx \epsilon_{n,l}\Big(\{V_i\}\Big).
\end{equation}
A potential with the postulated $3n+l$ degeneracy can then be found by
numerically optimizing a suitable initial potential with respect to a
cost function, which measures the deviation from the desired degeneracy.
By a straightforward application of the above strategy we have found
that the $V_i$'s for successive $i$'s oscillate wildly. 
For such potentials $\{V_i\}$ the approximation
(\ref{Ediscrete}) is clearly invalid. Thus we have imposed a
constraint on the $V_i$'s to keep the potential smooth. Furthermore we
have made sure that the number of electrons the potential can hold
for a given Fermi energy does not vary too much.

We have performed such optimizations starting from Woods-Saxon
potentials
\begin{equation}
  V(r) = {-V_0 \over 1+\exp\left({r-R_0\over a}\right)}
\end{equation}
with sodium-like parameters ($V_0=0.46\,Ry$ and $a=0.94\,a_0$) for
varying cluster sizes $R_0$. For the discretization we have used a mesh of
step width $\Delta r \approx r_s/40$.

\noindent
\begin{minipage}{3.375in}
\begin{figure}
  \centerline{\epsfxsize=3.37in \epsffile{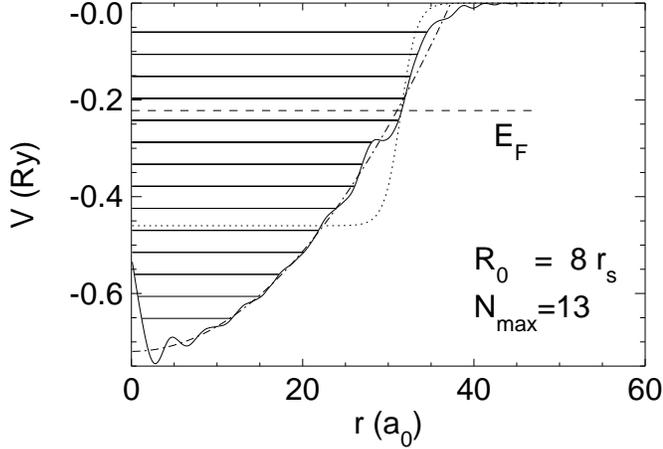}}
  \vspace{1ex}
  \caption[]{\label{harmopt}
             Result of an optimization run for the quantum number $2n+l$.
             The optimized potential (full line) along with the energy 
             levels (horizontal lines) are plotted. The computation was 
             started from a Woods-Saxon potential of width $R_0=8\,r_s$, 
             with $r_s=3.93\,a_0$ (dotted line). 
             The quadratic potential (dashed-dotted line) exhibiting the 
             exact $2n+l$-degeneracy is given for comparison.
             }
\end{figure}
\end{minipage}

\vspace{2ex}
\noindent
A cost function has measured
the non-degeneracy of the energy levels, the smoothness of the
potential and the variation in the number of electrons that fit into
the potential for a given Fermi energy. To find the non-degeneracy we have
determined for each quantum number $N_3=3n+l$ the average energy
\begin{equation}
  \bar{\epsilon}_{N_3} = 
  {\sum_{3n+l=N_3} 2(2l+1)\;\epsilon_{n,l} \over \sum_{3n+l=N_3} 2(2l+1)}.
\end{equation}
The deviation of the eigenstates from the desired degeneracy was
then defined as 
\begin{equation}\label{nondegen}
  \sum_{N_3} 
  {\sum_{3n+l=N_3} 2(2l+1)\;(\epsilon_{n,l}-\bar{\epsilon}_{N_3})^2 
   \over \sum_{3n+l=N_3} 2(2l+1)}.
\end{equation}
As a measure of the roughness of the potential represented by the
$V_i$ we have used the expression
\begin{equation}
  \sum_i (V_{i-1}-2V_i+V_{i+1})^2 ,
\end{equation} 
which disfavors sharp bends in $V(r)$. The number of electrons
fitting into the potential was estimated by the Thomas-Fermi
expression
\begin{equation}
  N_{el} = {4\over3\pi} \int_0^{r_{out}} dr\,r^2\;
            \left({2m\over\hbar^2}\Big(E_F-V(r)\Big)\right)^{3/2},
\end{equation}
where $r_{out}$ is the classical turning point for the potential $V(r)$.

Since the Woods-Saxon potential has only a finite number of bound
states, the summation (\ref{nondegen}) has to end at some maximal
$N$, which we will denote by $N_{max}$. 

To demonstrate that with the above measures we indeed can find potentials 
with a given degeneracy, we have 

\noindent
\begin{minipage}{3.375in}
\begin{figure}
  \centerline{\epsfxsize=3.37in \epsffile{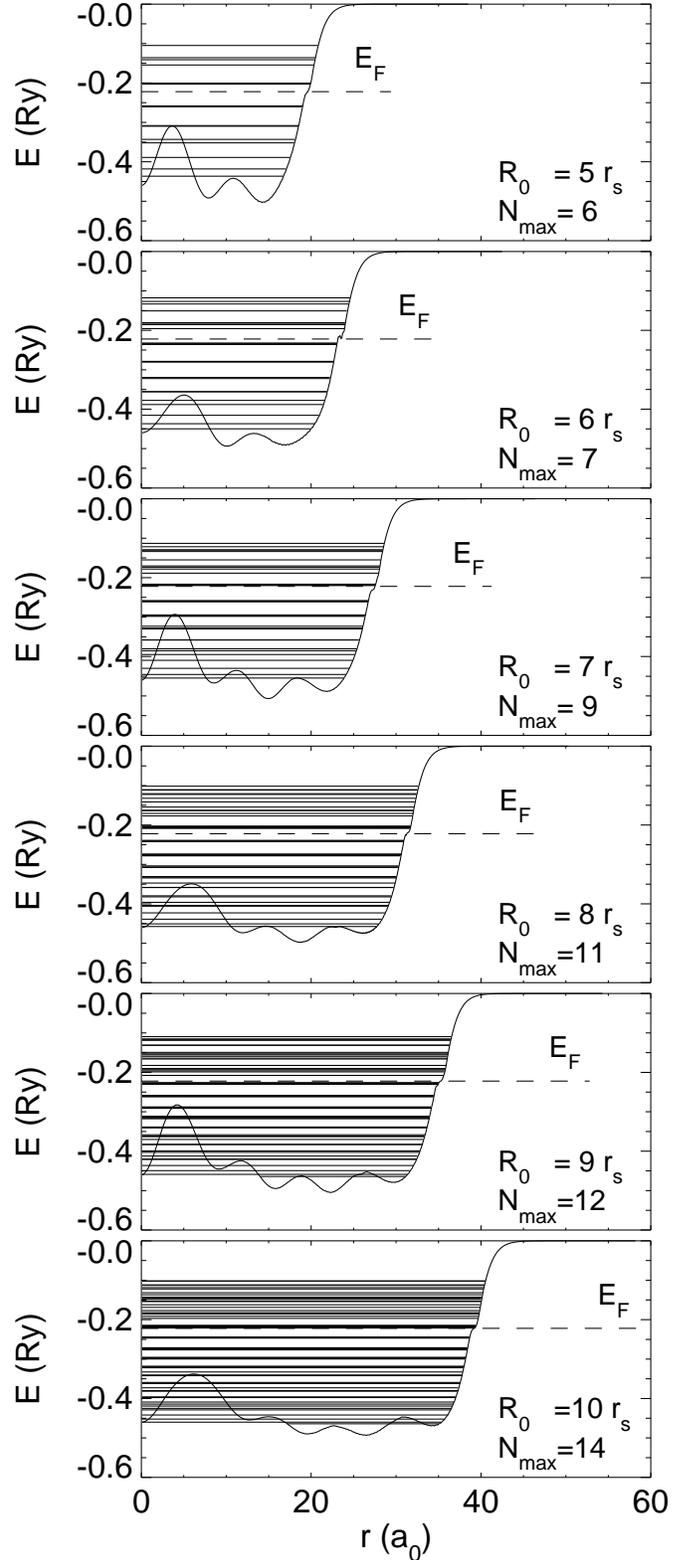}}
  \vspace{1ex}
  \caption[]{\label{potpanel}
    Potentials for increasing cluster radius $R_0$, optimized for degeneracies
    according to the quantum number $3n+l$. The optimization was started from
    Woods-Saxon potentials with sodium-like parameters. $r_s=3.93\,a_0$ is the
    Wigner-Seitz radius of Na. The lines indicate the position of the energy
    levels.}
\end{figure}
\end{minipage}
\end{multicols}

\begin{figure*}[tbh]
  \centerline{\epsfxsize=6.75in \epsffile{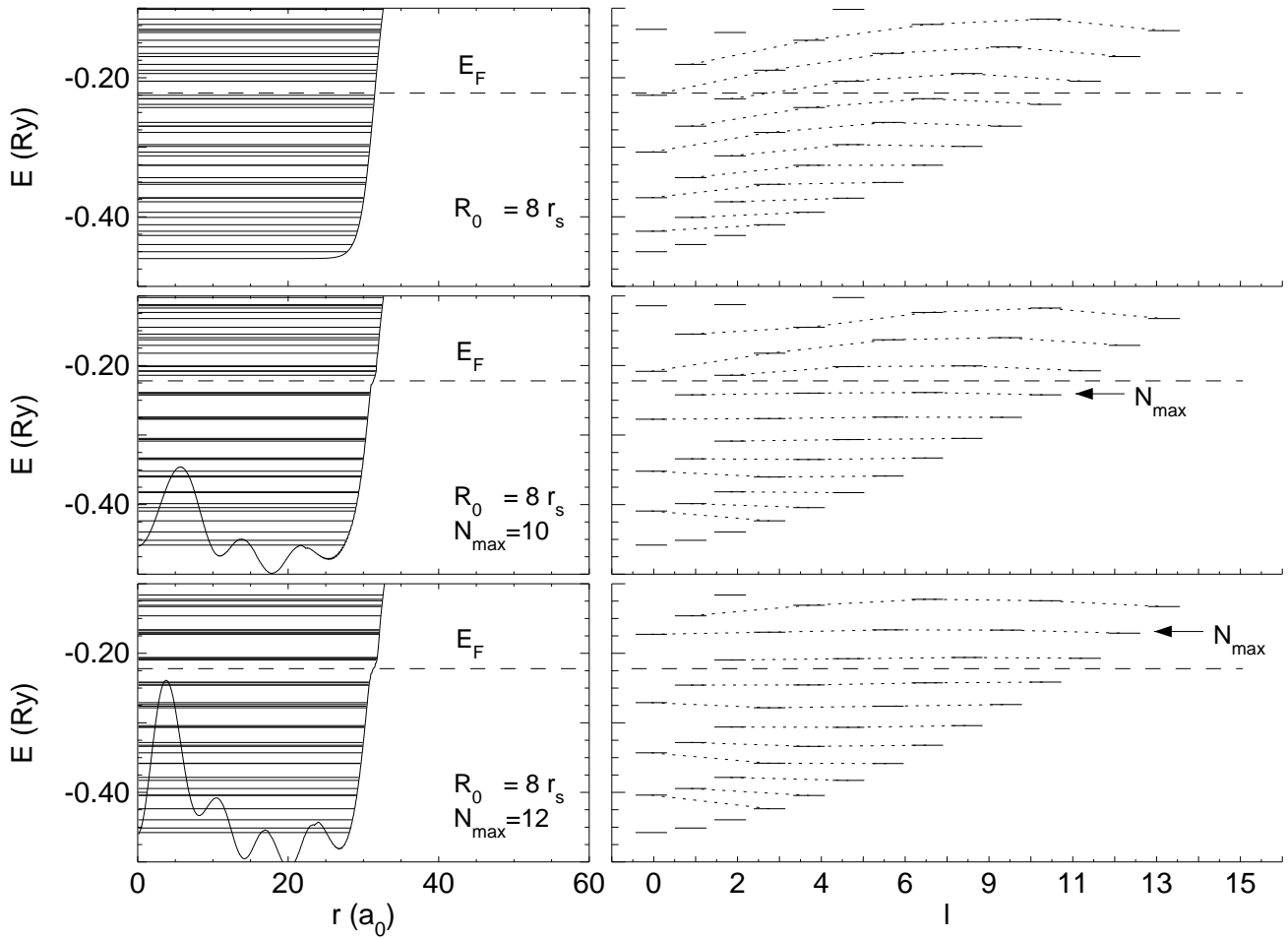}}
  \vspace{2ex}
  \caption[]{\label{levpanel}
    Optimization of cluster potentials for degeneracies according to the 
    quantum number $3n+l$. On the left, potentials are plotted along with the
    energy levels. The grouping of levels is shown
    on the right, where the eigenenergies $\epsilon_{n,l}$ are plotted over 
    the corresponding angular momenta $l$. Levels having the same quantum
    number $3n+l$ are connected by a dotted line. 
    The plots at the top show the spectrum of the Woods-Saxon potential,
    which has been used as initial potential for the optimization runs shown in
    the plots below. These runs only differ in the number $N_{max}$ of shells
    included in the cost function. }
\end{figure*} 

\vspace{1ex}
\begin{multicols}{2}
\noindent
tested our program on the quantum number $2n+l$.
We start from the sodium-like Woods-Saxon potential that we also use
for the $3n+l$-optimizations. Although this initial potential is much
closer to a $3n+l$ than to a $2n+l$-degeneracy, it turns out that
during optimization the energy levels are very easily arranged into
equidistant shells. For all choices of $N_{max}$ nearly
perfect degeneracy is achieved for the shells with $N<N_{max}$. Interestingly,
even the levels which are excluded from the cost function (i.e.\ those above
$N_{max}$) tend to form degenerate shells. Figure \ref{harmopt}
shows the energy spectrum along with the potential resulting from the
optimization of a Woods-Saxon potential with $R_0=8\,r_s$. As can be
seen the optimized potential approximates well the quadratic potential
(indicated by the dash-dotted line) associated with the $2n+l$ degeneracy,
although it is very different from the initial potential (dotted
line).\cite{inefficient}

The results of the optimizations for the quantum number $3n+l$ are
quite different. They are shown in figure \ref{potpanel}. To make the
results for different cluster sizes comparable, we have chosen $N_{max}$ 
to correspond always to the shell closest to Fermi energy. It turns
out that for small clusters the energy levels can be forced to form
nearly degenerate shells according to the quantum number $3n+l$. The
corresponding potentials, however, exhibit large oscillations inside
the cluster. These seem to be fairly unrealistic, when compared to
those obtained form self-consistent calculations.\cite{GenzkenPRL}
For larger clusters the situation becomes even worse. Here, although
the optimized potentials oscillate strongly, degeneracy can merely be
achieved for the states with $3n+l$ just below $N_{max}$. In other words, 
only just below $E_F$ a grouping of almost degenerate energy levels can 
be observed. Above and further below $N_{max}$ there are no large gaps 
in the spectrum, and hence no shells.

The above results suggest that the outcome of a degeneracy optimization strongly
depends on $N_{max}$, indicating that there is no unique `cluster
potential' which could be approximated by the optimized potentials.
To demonstrate this dependence, we have performed calculations with
different values of $N_{max}$ for a fixed cluster size. The results are
shown in Fig.~\ref{levpanel}. At the top the Woods-Saxon potential,
which was used as starting point in the optimization runs, along with
its energy spectrum is plotted. On the left-hand side the potential is
shown as in Fig.~\ref{potpanel} (except that we restrict the energy
axis to a smaller range). To the right the eigenstates
$\epsilon_{n,l}$ are arranged according to their angular momenta $l$.
States belonging to the same shell $N_3=3n+l$ are connected by a
dotted curve. A shell $N_3$ is (near-)degenerate if this curve is a
horizontal line. The greater its slope, the worse the degeneracy.
Optimizing the potential for two different values of $N_{max}$ results
in the potentials shown in the lower part of Fig.~\ref{levpanel}.
These potentials are markedly different.  Not only the amplitude of
the oscillations, even the number of maxima in $V(r)$ changes with
$N_{max}$. As can be seen from the spectra on the right hand side the
dotted curves connecting states, which belong to the same shell
$N_3=3n+l$ are concave for $N_3 > N_{max}$ and convex for
$N_3<N_{max}$.  Comparing the spectra for $N_{max}=10$ and $12$, we
observe that there is a tradeoff between straightening out the dotted
lines above $N_3=10$ and bending those corresponding to smaller quantum 
numbers; i.e.\ it seems to be impossible to obtain the desired degeneracy for
{\em all} shells. Furthermore it turns out that the higher $N_{max}$ (i.e.\
the more states are included in the cost function), the stronger are
the oscillations in $V(r)$.

If a `cluster potential' having a shape similar to a Woods-Saxon
potential existed, then the optimized potentials would have
approximated that unique potential. Like in the optimizations for the
quantum number $2n+l$, the degeneracies would become the better the
larger $N_{max}$. This is clearly not what we have found in our calculations.
Instead, the optimized potentials strongly depend on $N_{max}$ and further
we have not found any potential whose complete spectrum is degenerate 
according to the quantum number $3n+l$.

\section{Conclusions}

By use of analytic methods as well as numerical calculations we have
argued that there is no `cluster problem' analogous to the exactly
solvable problems of the hydrogen atom and the harmonic oscillator. 
The sequence of
quantum numbers $n+l$ and $2n+l$ characterizing the latter systems thus
cannot be simply extended to quantum numbers $3n+l$. In this
sense it seems that the quantum numbers $n+l$ and $2n+l$ are unique.

In spite of this, the higher `quantum numbers' $\alpha n + \beta l$ do 
have some meaning. The analysis of the energy levels {\em near the Fermi energy}
found in self-consistent jellium calculations reveals that the electronic
shell and supershell structure mainly arises from an interplay of near 
degeneracies of the type $3n+l$ {\em and} $4n+l$.\cite{ToBePublished} 
As has been pointed out by Bohr and Mottelson,\cite{BoMo}
such near degeneracies of energy levels corresponding to numbers 
$\alpha n + \beta l$ are related to classical periodic orbits.
For example, triangular orbits are connected with $3n+l$-degeneracies, and
square orbits with $4n+l$. This simple correspondence is however lost in a
full semiclassical analysis. Nevertheless for typical cluster potentials 
a periodic orbit expansion\cite{gutzwiller70} shows that the electronic shell
and supershell structure is essentially captured by only considering the most 
important contributions which stem from the triangular and the square orbits. 
Using the simple picture of Bohr and Mottelson this is consistent with the
conjectured degeneracies as well as with those that can be found in jellium
calculations.

Thus, the structures observed in the mass spectra of warmed
metal clusters are intimately related with the quantities $3n+l$ and
$4n+l$. These numbers are, however, not analogous to the quantum numbers
known from the exactly solvable models.

\section*{Acknowledgments}
I am much indebted to O.\ Gunnarsson for his invaluable advice.
Helpful discussions with T.~P.~Martin are gratefully acknowledged.
B.~Farid carefully read the manuscript and made a number of valuable
suggestions.

\bibliographystyle{prsty_long}

\begin{thebibliography}{10}
\bibitem{nqm}
To avoid confusion, we note that the principal 
quantum number used in atomic physics is given by $n+l+1$.
\bibitem{inefficient}
Using a radial mesh of width $\Delta r=0.1\,a_0$, it takes about two days on 
an IBM RS/6000-520 to obtain the result shown in figure \ref{harmopt}. This 
inefficiency is largely due to our particular choice for the initial potential.
\bibitem{ToBePublished}
A detailed analysis of how the Kohn-Sham orbitals are populated in 
self-consistent jellium calculations will be published elsewhere.
\end{thebibliography}

\end{multicols}
\end{document}